\documentclass[aps,prd,superscriptaddress,preprint,tightenlines,nofootinbib,showpacs]{revtex4}

\usepackage{epsfig}
\usepackage{graphicx}
\usepackage{graphics}
\usepackage{dcolumn}
\usepackage{bm}
\usepackage{afterpage}

\newcommand{\etal}{{\it et al.}}

\begin{document}

\preprint{\tighten\vbox{
                        \hbox{\hfil CLNS 09/2057}
                         \hbox{\hfil CLEO-09-10}
}}
\title{\boldmath Study of the semileptonic decay $D_s^+\to f_0(980) e^+\nu$ and implications for $B_s^0\to J/\psi f_0$}

\author{K.~M.~Ecklund}
\affiliation{Rice University, Houston, Texas 77005, USA}
\author{Q.~He}
\author{J.~Insler}
\author{H.~Muramatsu}
\author{C.~S.~Park}
\author{E.~H.~Thorndike}
\author{F.~Yang}
\affiliation{University of Rochester, Rochester, New York 14627, USA}
\author{M.~Artuso}
\author{S.~Blusk}
\author{S.~Khalil}
\author{R.~Mountain}
\author{K.~Randrianarivony}
\author{T.~Skwarnicki}
\author{S.~Stone}
\author{J.~C.~Wang}
\author{L.~M.~Zhang}
\affiliation{Syracuse University, Syracuse, New York 13244, USA}
\author{G.~Bonvicini}
\author{D.~Cinabro}
\author{A.~Lincoln}
\author{M.~J.~Smith}
\author{P.~Zhou}
\author{J.~Zhu}
\affiliation{Wayne State University, Detroit, Michigan 48202, USA}
\author{P.~Naik}
\author{J.~Rademacker}
\affiliation{University of Bristol, Bristol BS8 1TL, UK}
\author{D.~M.~Asner}
\author{K.~W.~Edwards}
\author{J.~Reed}
\author{A.~N.~Robichaud}
\author{G.~Tatishvili}
\author{E.~J.~White}
\affiliation{Carleton University, Ottawa, Ontario, Canada K1S 5B6}
\author{R.~A.~Briere}
\author{H.~Vogel}
\affiliation{Carnegie Mellon University, Pittsburgh, Pennsylvania 15213, USA}
\author{P.~U.~E.~Onyisi}
\author{J.~L.~Rosner}
\affiliation{University of Chicago, Chicago, Illinois 60637, USA}
\author{J.~P.~Alexander}
\author{D.~G.~Cassel}
\author{R.~Ehrlich}
\author{L.~Fields}
\author{L.~Gibbons}
\author{S.~W.~Gray}
\author{D.~L.~Hartill}
\author{B.~K.~Heltsley}
\author{J.~M.~Hunt}
\author{D.~L.~Kreinick}
\author{V.~E.~Kuznetsov}
\author{J.~Ledoux}
\author{H.~Mahlke-Kr\"uger}
\author{J.~R.~Patterson}
\author{D.~Peterson}
\author{D.~Riley}
\author{A.~Ryd}
\author{A.~J.~Sadoff}
\author{X.~Shi}
\author{S.~Stroiney}
\author{W.~M.~Sun}
\affiliation{Cornell University, Ithaca, New York 14853, USA}
\author{J.~Yelton}
\affiliation{University of Florida, Gainesville, Florida 32611, USA}
\author{P.~Rubin}
\affiliation{George Mason University, Fairfax, Virginia 22030, USA}
\author{N.~Lowrey}
\author{S.~Mehrabyan}
\author{M.~Selen}
\author{J.~Wiss}
\affiliation{University of Illinois, Urbana-Champaign, Illinois 61801, USA}
\author{M.~Kornicer}
\author{R.~E.~Mitchell}
\author{M.~R.~Shepherd}
\author{C.~M.~Tarbert}
\affiliation{Indiana University, Bloomington, Indiana 47405, USA }
\author{D.~Besson}
\affiliation{University of Kansas, Lawrence, Kansas 66045, USA}
\author{T.~K.~Pedlar}
\author{J.~Xavier}
\affiliation{Luther College, Decorah, Iowa 52101, USA}
\author{D.~Cronin-Hennessy}
\author{K.~Y.~Gao}
\author{J.~Hietala}
\author{R.~Poling}
\author{P.~Zweber}
\affiliation{University of Minnesota, Minneapolis, Minnesota 55455, USA}
\author{S.~Dobbs}
\author{Z.~Metreveli}
\author{K.~K.~Seth}
\author{B.~J.~Y.~Tan}
\author{A.~Tomaradze}
\affiliation{Northwestern University, Evanston, Illinois 60208, USA}
\author{S.~Brisbane}
\author{J.~Libby}
\author{L.~Martin}
\author{A.~Powell}
\author{}
\author{C.~Thomas}
\author{G.~Wilkinson}
\affiliation{University of Oxford, Oxford OX1 3RH, UK}
\author{H.~Mendez}
\affiliation{University of Puerto Rico, Mayaguez, Puerto Rico 00681}
\author{J.~Y.~Ge}
\author{D.~H.~Miller}
\author{I.~P.~J.~Shipsey}
\author{B.~Xin}
\affiliation{Purdue University, West Lafayette, Indiana 47907, USA}
\author{G.~S.~Adams}
\author{D.~Hu}
\author{B.~Moziak}
\author{J.~Napolitano}
\affiliation{Rensselaer Polytechnic Institute, Troy, New York 12180, USA}
\collaboration{CLEO Collaboration}
\noaffiliation

\date{July 18, 2009}

\begin{abstract}
Using  $e^+e^-\to D_s^-D_s^{*+}$ and $D_s^{*-}D_s^{+}$ interactions
at 4170 MeV collected with the CLEO-c detector, we investigate the semileptonic decays
$D_s^+\to f_0(980) e^+ \nu$, and $D_s^+\to \phi e^+ \nu$.
By examining the decay rates as  functions of the four-momentum transfer squared, $q^2$, we measure the ratio
$\left[\frac{d{\cal B}}{dq^2}(D_s^+\to f_0(980) e^+ \nu){\cal B}(f_0\to \pi^+\pi^-)\right]/
\left[\frac{d{\cal B}}{dq^2}(D_s^+\to \phi e^+ \nu) {\cal B}(\phi\to K^+K^-)\right]$
at $q^2$ of zero to be (42$\pm$11)\%. This ratio has been
predicted to equal the rate ratio $\left[{\cal B}(B_s\to J/\psi f_0){\cal B}(f_0\to \pi^+\pi^-)\right]/\left[{\cal B}(B_s\to J/\psi \phi){\cal B}(\phi\to K^+K^-)\right]$, thus indicating that the
CP eigenstate $J/\psi f_0$ could be useful for measuring CP violation via $B_s$ mixing.  Assuming a simple pole model for the
 form factor $|f_+(q^2)|$ in the $f_0e^+\nu$ decay, we find a pole mass of
($1.7^{+4.5}_{-0.7}\pm 0.2$) GeV. We also
determine the $f_0$ mass and width as $(977^{+11}_{-9}\pm 1){\rm~MeV}$, and $(91^{+30}_{-22}\pm 3)~{\rm MeV}$, respectively.
In addition, we present updated results for
${\cal B}(D_s^+\to f_0(980) e^+ \nu){\cal B}(f_0\to \pi^+\pi^-)=(0.20\pm 0.03\pm 0.01)$\%, and ${\cal B}(D_s^+\to \phi e^+ \nu)=(2.36\pm 0.23\pm 0.13)\%$.  Assuming that the $f_0$  wavefunction is a combination of strange and non-strange quark-antiquark components, we use our measurement for ${\cal B}(D_s^+\to f_0(980) e^+ \nu)$ to extract a value of the mixing angle that we find consistent with $\mid\overline{s}s\rangle$ dominance, adding to the mystery as to why the $f_0$ decays predominantly to two pions rather than two kaons.
\end{abstract}

\maketitle \tighten


\section{Introduction}
\begin{figure}[htb]
\includegraphics[width=74mm]{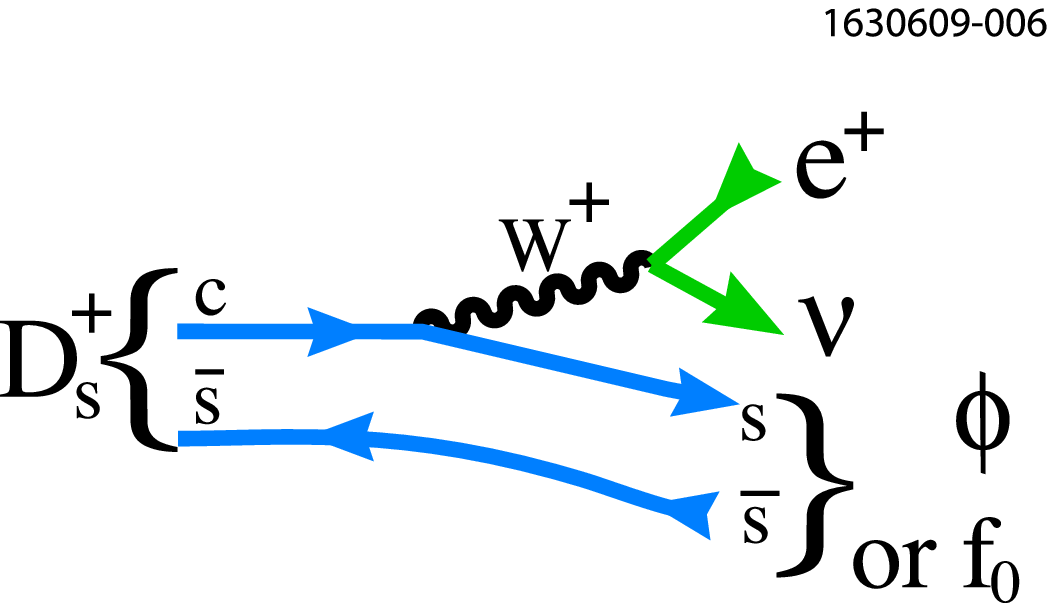}
\vspace{0.44mm}\caption{The Feynman diagram for semileptonic $D_s^+$ decay into a $\phi$ or $f_0(980)$ meson.} \label{Dssemi}
\end{figure}

In this article we present a study of semileptonic decay of the $D_s^+$ meson into $f_0(980)e^+\nu$
and also to $\phi e^+\nu$.
The $f_0(980)$ meson, a scalar, though well established experimentally, has a relatively uncertain
mass and width \cite{PDG}; in addition there have been claims that the quark content may be a mixture
of a traditional quark-antiquark with a four-quark system \cite{Joe}.  Semileptonic $D_s^+$ decays provide a pristine environment where the $f_0$ is produced by an isoscalar combination of $s$ and $\overline{s}$ quarks.
Evidence for $f_0(980)$ in semileptonic decays was seen by BaBar via the $f_0(980)\to K^+K^-$ channel, where interference was observed by an S-wave with the dominant P-wave $\phi$ decay \cite{BaBar-f0}. The first measurement of the $D_s^+\to f_0(980) e^+\nu$, $f_0\to\pi^+\pi^-$ product branching fraction was recently made by CLEO \cite{CLEO-semi}. The semileptonic decay diagram is shown in Fig.~\ref{Dssemi}. Here we investigate this mode using a data sample of 600 pb$^{-1}$, approximately double the original size. This larger sample is sufficient to allow us to determine the $f_0$ mass and width and measure the semileptonic decay form factor as a function of the invariant four-momentum transfer squared, $q^2$, between the $D_s$ and the $f_0$.

CP violation measurements in the $B_s$ system have concentrated on the final state $J/\psi\phi$, with $\phi\to K^+K^-$.
For a review see Ref.~\cite{Our-review}. Since this mode is not a CP eigenstate an angular analysis is required to separate the CP even and CP odd parts, and measure the CP violating phase $-2\beta_s$.

In what follows we will use a simple notation for the decay width of a
particle multiplied by the branching fraction for the decay of one of
its daughters. For example,
\begin{equation}
\Gamma(D_s^+\to \pi^+ f_0,~f_0\to \pi^+\pi^-)\equiv \Gamma(D_s^+\to \pi^+ f_0(980)){\cal{B}}(f_0(980)\to \pi^+\pi^-).
\end{equation}

Stone and Zhang \cite{SZ} have suggested that the  $J/\psi\phi$ mode may also contain an S-wave $K^+K^-$ system at the $\phi$ mass with a rate that could be as large as $\approx$5\% that of the $\phi$. This S-wave, if it should be significant, would require additional parameters in the fit to extract $-2\beta_s$. They also suggest that the final state $J/\psi f_0$ may be a useful alternative; since it is a CP-eigenstate angular analysis is not required. Stone and Zhang estimated the branching ratio $B_s\to J/\psi f_0$ assuming equality of the ratios
\begin{equation}
R_{f/\phi}\equiv\frac{\Gamma\left(B_s^0\to J/\psi f_0,~f_0\to\pi^+\pi^-\right)}{\Gamma\left(B_s^0\to J/\psi \phi,~\phi\to K^+K^-\right)}
=\frac{\Gamma\left(D_s^+\to f_0\pi^+,~f_0\to\pi^+\pi^-\right)}{\Gamma\left(D_s^+\to \phi\pi^+,~\phi\to K^+K^-\right)}
\approx (20{\rm-}30)\%~.
\end{equation}
The phase spaces for the $B_s^0$ and $D_s^+$ decays are quite similar, but the spin structure is not. In these $D_s^+$ decays both the $\phi$ and the $f_0$ are produced opposite a spin-0 pion, rather than a spin-1 $J/\psi$. This consideration prompted Stone and Zhang to suggest that measuring the ratio of decay widths of semileptonic $D_s^+$ decays containing either an $f_0$ or a $\phi$ at four-momentum transfers, $q^2$, equal to zero would give a superior prediction \cite{SZ}. Specifically they propose that
\begin{equation}
R_{f/\phi}= \frac{{\frac{d\Gamma}{dq^2}}(D_s^+\to f_0(980) e^+\nu,~f_0\to\pi^+\pi^-)\mid_{q^2=0}}{{\frac{d\Gamma}{dq^2}}(D_s^+\to \phi e^+\nu,~\phi\to K^+ K^-)\mid_{q^2=0}}~.
\end{equation}
The point $q^2$ equal to zero is chosen to maximize the allowed phase space in order to make
it as close as possible to that available in $B_s^0\to J/\psi \phi$ (or $f_0$) decay.
In this paper we will present measurements of this ratio,  the form factor in the $D_s^+\to f_0 e^+\nu$ channel,
 the $f_0$ mass and width, and update the previously published CLEO branching fractions for these two semileptonic decay modes \cite{CLEO-semi}.

\section{Experimental Method}
\subsection{Selection of $D_s$ Candidates}

The CLEO-c detector \cite{CLEODR} is equipped to measure the momenta
and directions of charged particles, identify them using specific
ionization ($dE/dx$) and Cherenkov light (RICH) \cite{RICH}, detect
photons and determine their directions and energies.

In this study we use 600 pb$^{-1}$ of data produced in $e^+e^-$
collisions using the Cornell Electron Storage Ring (CESR) and
recorded near a center-of-mass energy ($E_{\rm CM}$) of 4.170 GeV.
At this energy the $e^+e^-$ annihilation cross-section into
$D_s^-D_s^{*+}$ plus $D_s^{*-}D_s^{+}$ is approximately 1~nb \cite{poling}.

In this analysis we fully reconstruct a sample of $D_s^-$
in several ``tag" modes and then find candidate semileptonic
decays in this sample. Mention of any specific decay implies
the use of its charge-conjugate as well. The tag selection is
identical to that used in our $D_s^+\to \mu^+\nu$ paper, that
can be consulted for details \cite{Dstomunu}. Briefly, we select
candidates on the basis of their beam-constrained invariant mass. Then we detect
an additional photon candidate from the $D_s^{*}$ decay, and construct the missing
mass squared, MM$^{*2}$ recoiling against the photon and the $D_s^-$ tag
\begin{equation}
\label{eq:mmss} {\rm MM}^{*2}=\left(E_{\rm
CM}-E_{D_s}-E_{\gamma}\right)^2- \left({\bf p}_{\rm
CM}-{\bf p}_{D_s}-{\bf p}_{\gamma}\right)^2,
\end{equation}
where $E_{\rm CM}$ (${\bf p}_{\rm CM}$) is the
center-of-mass energy (momentum), $E_{D_s}$
(${\bf p}_{D_s}$) is the energy (momentum) of the fully
reconstructed $D_s^-$ tag, and $E_{\gamma}$
(${\bf p}_{\gamma}$) is the energy (momentum) of the
additional photon. In performing this calculation we use a kinematic
fit that constrains the decay products of the $D_s^-$ to the known
$D_s$ mass and conserves overall momentum and energy. All photon
candidates in the event are used, except for those that are decay
products of the $D_s^-$ tag candidate.
 Regardless of whether
or not the photon forms a $D_s^*$ with the tag, for real $D_s^*D_s$
events MM$^{*2}$ should peak at the $D_s^{+}$ mass-squared.

\begin{figure}[hbt]
\centering
\includegraphics[width=6in]{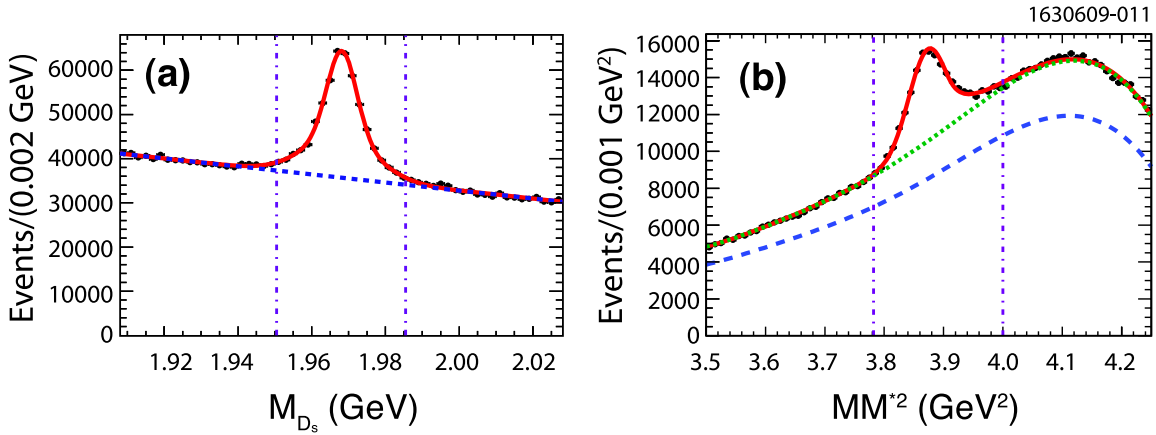}
\vspace{2mm}
\caption{(a) Invariant mass of $D_s^-$ candidates
summed over all decay modes and fit to a two-Gaussian signal shape
plus a straight line for the background. The vertical dot-dashed
lines indicate the $\pm$17.5 MeV definition of the signal region. (b)
The MM$^{*2}$ distribution summed over all modes. The
curves are fits to the number of signal events using the Crystal Ball function
and two 5th order
Chebychev background functions; the dashed curve shows the
background from fake $D_s^-$ tags, while the dotted curve in (b) shows
the sum of the backgrounds from multiple photon combinations and
fake $D_s^-$ tags. The vertical dashed lines show
the region of events selected for further analysis.
 } \label{mass-mm2-all}
\end{figure}

We list the number of signal events in each mode in
Table~\ref{tab:Ntags} by finding the number of events within
$\pm$17.5 MeV of the $D_s$ mass. For ease of further analysis we sum all
tag modes together, as shown in Fig.~\ref{mass-mm2-all}(a).

\begin{table}[htb]
\begin{center}
\caption{Tagging modes and numbers of signal and background events,
within $\pm$17.5 MeV of the $D_s^-$ mass for each mode, determined
from two-Gaussian function fits to the invariant mass plots, and the number
of tags in each mode including the $\gamma$ from the $D_s^*\to\gamma
D_s$ transition, within an interval $3.782 < {\rm MM}^{*2}<4.0$ GeV$^2$, as
determined from fits of the MM$^{*2}$ distributions (see text) to a
signal Crystal Ball function (see text) and two 5th order Chebychev
background polynomial functions.\label{tab:Ntags}}
\begin{tabular}{lcrcr}
 \hline\hline
    Mode  & \multicolumn{2}{c}{Invariant Mass}& \multicolumn{2}{c}{MM$^{*2}$}\\
    &  Signal & Background & Signal & Background \\\hline
$K^+K^-\pi^- $ & 26534$\pm$274 & 25122 &16087$\pm$373 &39563\\
$K_S K^-$ & 6383$\pm$121 & 3501 & 4215$\pm$228&6297\\
$\eta\pi^-$; $\eta\to\gamma\gamma$ & $2993\pm156$  &
5050&2005$\pm$145 &5016\\
$\eta'\pi^-$; $\eta'\to\pi^+\pi^-\eta$, $\eta\to\gamma\gamma$
& 2293$ \pm $82  &531 &1647$\pm$131 &1565 \\
$K^+K^-\pi^-\pi^0$ & 11649$ \pm $754  &78588&6441$\pm$471&89284\\
$\pi^+\pi^-\pi^-$ & 7374$ \pm $303  & 60321 & 5014$\pm$402& 43286\\
$K^{*-}K^{*0}$; $K^{*-}\to K_S^0\pi^-$, ${K}^{*0}\to K^+\pi^-$ &
4037$\pm$160& 10568&2352$\pm$176 & 12088\\
$\eta\rho^-$; $\eta\to\gamma\gamma$, $\rho^-\to \pi^-\pi^0$
& 5700$ \pm $281  &24444 & 3295$\pm$425 & 24114\\
$\eta'\pi^-$; $\eta'\to\rho^0\gamma$,
& 3551$ \pm $202  &19841 &2802$\pm$227 &17006 \\
\hline
Sum &  $70514\pm 963 $ &227966 & 43859$\pm$936&238218\\
\hline\hline
\end{tabular}
\end{center}
\end{table}

The MM$^{*2}$ distributions for events in the $D_s^-$ invariant mass
signal region ($\pm$17.5 MeV from the $D_s$ mass) are shown in
Fig.~\ref{mass-mm2-all}(b). In order to find the number of tags used for
further analysis we perform a two-dimensional binned maximum likelihood fit of the MM$^{*2}$
distribution and the invariant mass distribution in the interval $\pm$60 MeV from the
$D_s$ mass and $3.5 < {\rm MM}^{*2} <4.25$~GeV$^2$. The background has two components, both
described by 5th order Chebyshev polynomials in ${\rm MM}^{*2}$; the first comes from
the background under the invariant mass peak, defined by the
sidebands, and the second is due to multiple photon combinations. In
both cases we allow the parameters to float.

 We find a
total of 43859$\pm$936$\pm$877 events within the interval
$3.782 < {\rm MM}^{*2}<4.0$ GeV$^2$ and having an invariant mass within $\pm$17.5 MeV of the $D_s$ mass,
where the total number of events is the sum of the yields from the fits to
each mode as shown in Table~\ref{tab:Ntags}. The first uncertainty in the total is
statistical and the second is systematic.

\subsection{Signal Reconstruction}
\label{sec:bfqsq}
We next describe the reconstruction of $D_s^+\to f_0 e^+ \nu,~f_0\to \pi^+\pi^-$, and
also $D_s^+\to \phi e^+ \nu,~\phi\to K^+ K^-$. We select events
within the MM$^{*2}$ region shown in Fig.~\ref{mass-mm2-all}(b) for
further analysis. We note that the limits are rather wide. We use
this selection because the background in
the signal side is rather small and the errors are minimized by
taking as many tags as possible.

Candidate events are selected that contain a charged track of
opposite sign to the $D_s^-$ tag that is positively identified as
a positron. Electrons and positrons are identified on
the basis of a likelihood ratio constructed from three inputs: the ratio between the energy
deposited in the calorimeter and the momentum measured in the tracking system, the specific
ionization $dE/dx$ measured in the drift chamber, and RICH information \cite{Coan}. Our selection
efficiency averages 0.95 in the momentum region 0.3-1.0 GeV, and 0.71 in the region 0.2-0.3
GeV. The average fractions of charged pions and kaons incorrectly identified as positrons averaged
over the relevant momentum range are approximately 0.1\%. We also require an additional pair of tracks
with opposite charge that are both identified as pions or kaons using the $dE/dx$ and RICH systems.

Since we are searching for events containing a single missing
neutrino the missing mass squared for the $f_0$ mode, MM$^2$,  evaluated by taking into
account the observed $e^+,~\pi^+,~\pi^-$, $D_s^-$, and $\gamma$ should peak at
zero; the MM$^2$ is computed as

\begin{equation}
\label{eq:mm2} {\rm MM}^2=\left(E_{\rm
CM}-E_{D_s}-E_{\gamma}-E_{e}-E_{\pi^+}-E_{\pi^-}\right)^2
           -\left({\bf p}_{\rm CM}-{\bf p}_{D_s}
           -{\bf p}_{\gamma}
           -{\bf p}_{e}-{\bf p}_{\pi^+}-{\bf p}_{\pi^-}
           \right)^2,
\end{equation}
where $E_{e}$ (${\bf p}_{e}$) are the energy
(momentum) of the candidate positron, $E_{\pi}$ (${\bf p}_{\pi}$)
are the energy
(momenta) of the candidate pions, and all other variables are the
same as defined in Eq.~(\ref{eq:mmss}). A similar equation applies for the $\phi$ mode
with the pions replaced by kaons.

We also make use of a set of kinematical constraints and fit each
event to two hypotheses, one of which is that the $D_s^-$ tag is the
daughter of a $D_s^{*-}$, and the other that the $D_s^{*+}$ decays
into $\gamma D_s^+$ with the $D_s^+$ subsequently decaying into either
$\pi^+\pi^- e^+\nu$, or $K^+ K^- e^+\nu$. The kinematical constraints, in the $e^+e^-$ center-of-mass
frame, are
\begin{eqnarray}
\label{eq:constr}
&&{\bf p}_{D_s}+{\bf p}_{D_s^*}=0,\\\nonumber
&&E_{\rm CM}=E_{D_s}+E_{D_s^*},\\\nonumber
&&E_{D_s^*}=\frac{E_{\rm
CM}}{2}+\frac{M_{D_s^*}^2-M_{D_s}^2}{2E_{\rm CM}}{\rm~or~}
E_{D_s}=\frac{E_{\rm CM}}{2}-\frac{M_{D_s^*}^2-M_{D_s}^2}{2E_{\rm
CM}},~{\rm and}\\\nonumber
&&M_{D_s^*}-M_{D_s}=143.8 {\rm ~MeV}.
\end{eqnarray}
In addition, we constrain the invariant mass of the $D_s^-$ tag to
the known $D_s$ mass. This gives us a total of seven constraints. The
missing neutrino four-vector needs to be determined, so we are left
with a three-constraint fit. We perform an iterative fit that
minimizes $\chi^2$. As we do not want to be subject to systematic
uncertainties that depend on understanding the absolute scale of the
errors, we do not make a $\chi^2$ cut but simply choose the photon
and the decay sequence in each event with the minimum $\chi^2$.

\begin{figure}[hbt]
\centering
\includegraphics[width=6in]{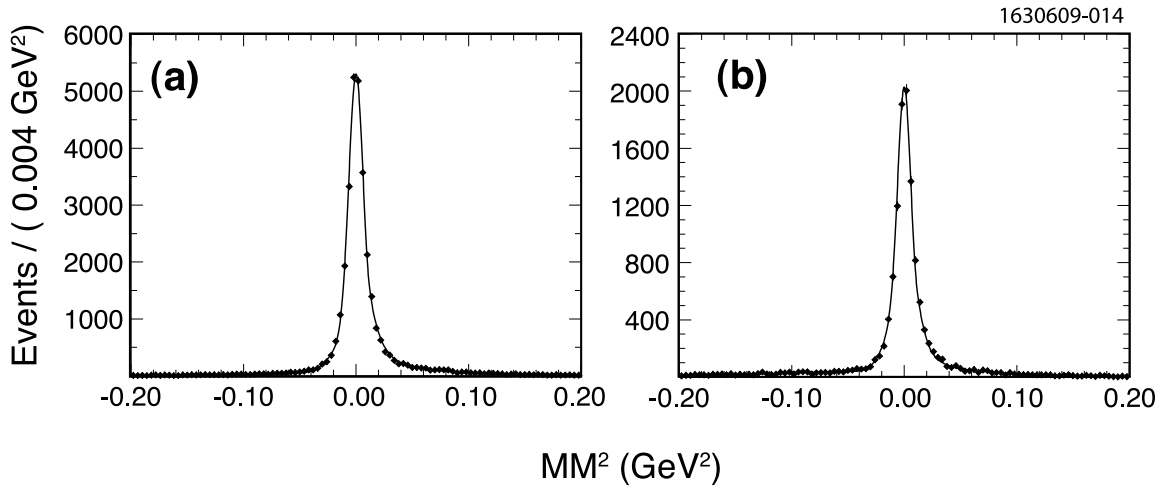}
\caption{MM$^2$ distributions from Monte Carlo simulation for
(a) $D_s^+\to f_0 e^+\nu$ and (b) $D_s^+\to \phi e^+\nu$. The curves are the sum of two CB functions with means fixed at zero, $n$ fixed at 1.8 and $\alpha$ values constrained to be equal, and a single Gaussian shape. (The parameters are listed in Table~\ref{tab:fit-params}.)} \label{mc-munu-res}
\end{figure}

We model the MM$^2$ signal distributions for both  $f_0 e^+\nu$  and $\phi e^+\nu$ final states as the sum of two Crystal Ball (CB) functions centered at zero \cite{CBL} plus a wide Gaussian shape that serves to model the tails. The Monte Carlo simulations are
shown in Fig.~\ref{mc-munu-res}. The results are summarized in Table~\ref{tab:fit-params}.

\begin{table}[htb]
\begin{center}
\caption{Parameters of signal fits to two CB functions plus a Gaussian shape; ``${\cal F}$" indicates the
fraction of the yield in each fit component. The parameter $n$ used in
both CB functions is fixed at 1.8. The parameter $\alpha$
is kept the same for both CB functions. The mean and width of the Gaussian is kept the
same for both decay modes. All means and $\sigma$'s are in units of GeV$^2$. \label{tab:fit-params}}
\begin{tabular}{lcccccccc}
 \hline\hline
 & \multicolumn{3}{c}{1$^{\rm st}$ CB function}&\multicolumn{2}{c}{2$^{\rm nd}$ CB function}&\multicolumn{3}{c}{Gaussian}\\
 &  ${\cal F}$(\%) &  $\sigma_1$ &$\alpha$ & ${\cal F}$(\%) & $\sigma_2$ &  ${\cal F}$(\%) & mean & $\sigma$  \\\hline
 $\pi^+\pi^-$ & 63 & $(6.30\pm 0.15)\times 10^{-3}$& 1.417$\pm$0.014 & 30 &  $(1.67\pm 0.07)\times 10^{-2}$ & ~7 & -0.0253 & 0.0992\\
$K^+K^-$ & 45 & $(5.61\pm 0.35)\times 10^{-3}$& 1.610$\pm$0.003 & 37 &  $(1.32\pm 0.09)\times 10^{-2}$ & 18 & -0.0253& 0.0992\\
\hline\hline
\end{tabular}
\end{center}
\end{table}

\begin{figure}[htbp]
 \vskip 0.00cm
\centerline{\epsfxsize=6.0in \epsffile{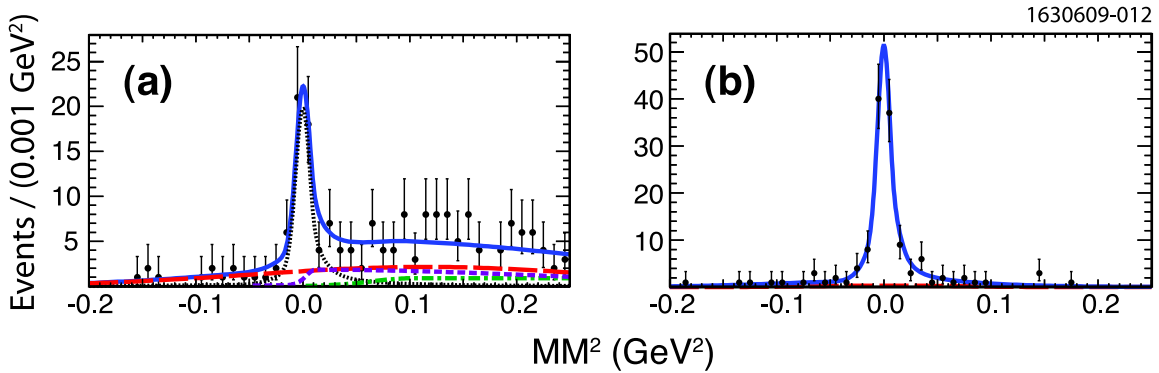}}
 \caption{(a) The MM$^2$ distribution for $\pi^+\pi^- e^+\nu$. The dotted curve shows the signal, the long-dashed distributions are from sidebands of the $D_s^-$ candidate invariant mass distributions while the shorter dashed curve in (a) shows the $\eta'\to\pi^+\pi^-\gamma$ background level, while the dashed-dot curve shows the (small) $\pi^+\pi^-\pi^0$ background from $\phi e^+\nu$. The solid curve shows the total. (b) The MM$^2$ distribution for $K^+ K^- e^+\nu$, where the total and the small side-band background is shown.  }
 \label{mm2-both-all}
 \end{figure}

We proceed by performing a simultaneous fit to the $D_s^-$ invariant mass, using a mass range $\pm$70 MeV from the
nominal mass, and the MM$^2$ for $\pi^+\pi^- e^+\nu$ and $K^+ K^- e^+\nu$ to the Monte Carlo generated fitting functions letting only the normalizations float.\footnote{This is the same procedure as used in our $D_s^+\to \mu^+\nu$ analysis.} The resulting MM$^2$ distributions are shown in Fig.~\ref{mm2-both-all}.  Here and in subsequent analyses we require that the  $\pi^+\pi^-$ invariant mass is above 0.6 GeV, in order to eliminate the $K_S^0 e^+\nu$ channel, and limit the $K^+K^-$ invariant mass to be below 1.08 GeV.

Our next step is to determine the range of MM$^2$ to select for further analysis. Thus, we consider the backgrounds in our MM$^2$ sample.
In general the background arises from two sources: one from real
$D_s^+$ decays and the other from the background under the
single-tag signal peaks (fake $D_s^-$). For $\phi e^+\nu$ only the background from fake $D_s^-$ is significant. For $f_0 e^+ \nu$
we show in Fig.~\ref{mm2-both-all}(a) the background from the invariant mass sidebands, suitably scaled. We also find that there are
two sources of real $D_s^+$ background, namely $\eta' e^+\nu$ decays where the $\eta' \to\rho\gamma$, and a very small component of
$\phi e^+\nu,$ $\phi\to\pi^+\pi^-\pi^0$ when the $\pi^0$ is ignored.
We also show these backgrounds as a function of MM$^2$. We use a second order polynomial to parameterize the fake $D_s^-$ background,
and the other two backgrounds are parameterized by the sum of
two bifurcated Gaussian shapes determined by simulating the background processes. (A bifurcated Gaussian shape has a different widths below and above the
mean.)  We are able to fix the $\pi^+\pi^-\pi^0 e^+ \nu$ rate as 0.31\%,
using our previous measurement of $D_s^+\to\phi e^+\nu$, but allow the $\pi^+\pi^-\gamma e^+ \nu$ yield to float in the fit as our measured branching fraction is only accurate to $\pm$35\% \cite{CLEO-semi}.

For further analysis we
restrict ourselves to the interval $-0.04<$ MM$^2$ $<$ 0.04 GeV$^2$,  which is 88.4\% efficient on $f_0 e^+ \nu$ and is 84.4\% efficient on $\phi e^+ \nu$. In these regions we find 42.9$\pm$6.7 $\pi^+\pi^- e^+\nu$ signal events,
7.7$\pm$2.0 $\eta' e^+\nu$ events, 0.5 $\phi e^+\nu$ events (fixed), and 13.4$\pm$0.9 fake $D_s^-$ background. There are 107.0$\pm$9.8 $K^+K^- e^+\nu$ events of which 2.5$\pm$0.7 are fake $D_s^-$ background.

\subsection{\boldmath $K^+ K^-$ and $\pi^+\pi^-$ Invariant Mass Distributions}
\label{sec-IMD}
\begin{figure}[htbp]
 \vskip 0.00cm
\centerline{\epsfxsize=6.0in \epsffile{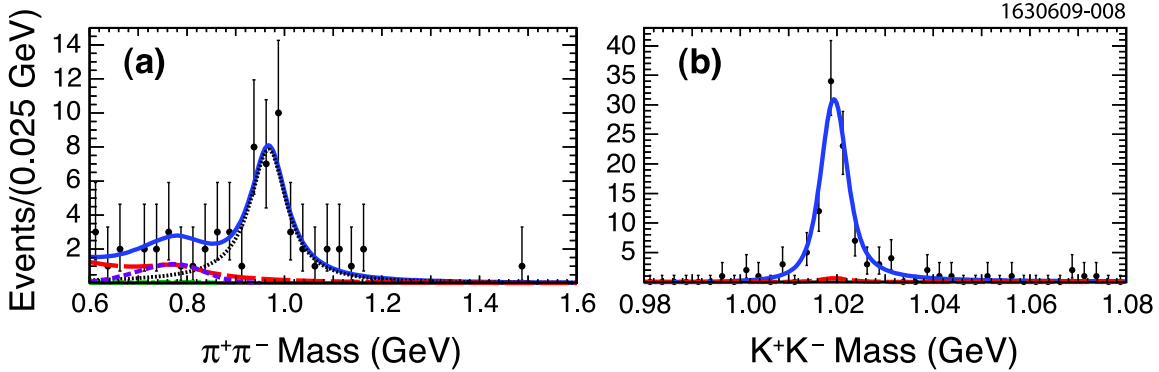}}
 \caption{Invariant mass distributions for (a) $\pi^+\pi^-$ and  (b) $K^+ K^-$ for semileptonic decays..
 The dotted curve shows the signal, the long-dashed distributions are from sidebands of the $D_s^-$ candidate invariant mass distributions while the shorter dashed curve in (a) shows the $\eta'\to\pi^+\pi^-\gamma$ background level, and the dashed-dot curve in (a) shows the (small) $\pi^+\pi^-\pi^0$ background from $\phi e^+\nu$. The solid curves show the totals.}
 \label{f0mass}
 \end{figure}

We next include the di-hadron mass as part of the joint fit to the $D_s^-$ mass, and MM$^2$.
Figure~\ref{f0mass} shows the $\pi^+\pi^-$ invariant mass distribution for
$\pi^+\pi^- e^+\nu$ events, and the $K^+K^-$ invariant mass distribution in our selected sample of $K^+ K^- e^+\nu$ events, within the interval  $-0.04<$ MM$^2$ $<$ 0.04 GeV$^2$. Both distributions are dominated by a single di-hadron resonance, either the $f_0(980)$ in the $\pi^+\pi^-$ mode or the $\phi$ in the $K^+K^-$ mode. The backgrounds are included in the fit in the same manner as for the MM$^2$ distributions.  We next find the size of the $f_0$ and $\phi$ signals.

For the $K^+ K^- e^+\nu$, the signal fitting function consists of the sum of a P-wave Breit-Wigner shape whose mass and
width are fixed to the values reported by the Particle Data Group (PDG) \cite{PDG} convoluted with our Gaussian experimental resolution, and a fixed background function determined from the $D_s^-$ sidebands described by the same $\phi$ signal shape added to a wide component that is exponentially decreasing with mass.

 For the $\pi^+\pi^-$ mass distribution we include three separate shapes to fit the backgrounds. (1) The $\pi^+\pi^-\gamma$ distribution is described by a bifurcated-Gaussian function; (2) the $\pi^+\pi^-\pi^0$ distribution is described by a bifurcated-Gaussian shape plus a Breit-Wigner shape describing the $\rho$, here  our knowledge of the $\phi e^+\nu$ decay rate allows us to fix the magnitude of this component; (3) the shape describing non-$D_s^-$ background is an exponential plus a Breit-Wigner function describing the $\rho$,  the size of this component is determined from the sidebands of the $D_s^-$ signal.

The mass and width of the $f_0(980)$ are not well determined. The PDG makes an estimate of the mass of (980$\pm$10) MeV, but does not explicitly average
any measurements. They also state that the width is very model dependent and give a range of 40-100 MeV, again not forming an average \cite{PDG}.
Here we provide measurements in an extremely clean environment. We fit the $\pi^+\pi^-$ mass distribution, $\sqrt{s}\equiv M_{\pi\pi}$ with a Breit-Wigner type resonance form with the threshold behavior suggested in \cite{DFNN}

\begin{equation}
\label{eq:relBW}
P(M_{\pi\pi}^2)= \frac{M_{f_0}\Gamma\left(s\right)}
{\left(s-M_{f_0}^2\right)^2+M_{f_0}^2\Gamma^2\left(s\right)},
\end{equation}
where $M_{f_0}$ denotes the $f_0$ mass and $\Gamma\left(s\right)$ the width
which is a function of the mass having the explicit dependence
\begin{equation}
\label{eq:relBW2}
\Gamma\left(s\right)=\Gamma_0
\frac{\sqrt{\left(s-4M^2_{\pi}\right)}}
{\sqrt{\left(M_{f_0}^2-4M^2_{\pi}\right)}}\frac{M_{f_0}}{M_{\pi\pi}},
\end{equation}
where $M_{\pi}$ is the charged pion mass, and $\Gamma_0$ the width in the limit
where the pion becomes massless.

We find that
$M_{f_0}=(968\pm 9){\rm~MeV}$, and $\Gamma_0=(92^{+28}_{-21})~{\rm MeV}$.
Later in this paper we will slightly refine these values of the mass and width, which we use for subsequent analysis. We note that performing this fit using a normal Breit-Wigner function gives $M_{f_0}=(966\pm 9){\rm~MeV},~\Gamma=(89^{+26}_{-20})~{\rm MeV}.$

\section{Branching Fractions and $q^2$ Distributions}
\subsection{Main Experimental Results}
The main aims of this paper are to provide the branching fraction and $q^2$ dependence for the $f_0 e^+ \nu$ mode and
to measure the relative rates at $q^2=0$ of the $f_0$ and $\phi$ modes. The BaBar collaboration previously measured
${\cal B}(D_s^+\to\phi e^+ \nu) = (2.61\pm 0.03 \pm 0.08 \pm 0.15)$\%,\footnote{The first error is statistical,
the second experimental systematic, and the third due to the absolute $D_s^+$ branching fraction scale.} and the form factors for this decay
given in the Appendix. They also found that the S-wave contribution to the $K^+ K^- e^+ \nu$ rate is a small fraction $(0.22^{+0.12}_{-0.08})\%$
of the total rate in the mass interval between 1.01 and 1.03 GeV \cite{BaBar-f0}.

In general, for a decay of a  pseudoscalar $D_s^+$ to a scalar meson such as the $f_0(980)$, the only Lorentz invariants in the problem are the invariant square $P^2$ of
the summed four-momentum of the $D_s$ and $f_0$, and the invariant square $q^2$ of the four-momentum transfer between the $D_s$ and the $f_0$. The hadronic current
describing the decay can be expressed as
\begin{equation}
\langle f_0\mid \overline{s}\gamma^{\mu}(1-\gamma^5)c\mid D_s\rangle = f_+(q^2)P^{\mu}+f_-(q^2)q^{\mu}~,
\end{equation}
where $f_{\pm}(q^2)$ are arbitrary form factors. Since this transition is between a pseudoscalar $D_s^+$ and a scalar $f_0$,
only the axial-vector part of the current contributes. The term containing $f_-(q^2)$ gets multiplied by the positron mass squared,
and becomes vanishingly small \cite{Artuso-semi}.

The decay rate can be written as
\begin{equation}
\label{eq:f0q2}
\frac{d\Gamma\left(D_s^+\to f_0 e^+\nu\right)}{dq^2}
=\frac{G_F^2\left|V_{cs}\right|^2}{24\pi^3}p^3_{f_0}\left|f_+(q^2)\right|^2~,
\end{equation}
where $G_F$ is the Fermi constant, $p_{f_0}$ is the magnitude of the three-momentum of the $f_0$ in the $D_s^+$ rest frame, and $|V_{cs}|$ is a Cabibbo-Kobayshi-Maskawa matrix element equal to $\approx0.97$ \cite{PDG}.
Therefore, we can use our data to determine $|f_+(q^2)|$.

Equation~\ref{eq:f0q2} is strictly correct only for zero width. It can be modified, however, to correct for the finite width mesons as described by Isgur~\etal~\cite{ISGW}. Here we integrate the product of  $P(s)$ as defined in Eq.~\ref{eq:relBW} with the right-hand side of Eq.~\ref{eq:f0q2} over $dM_{\pi\pi}$.

The form factors for  $\phi e^+ \nu$ are somewhat more complicated as the hadronic transition involves a transformation from a pseudoscalar $D_s$ to a vector $\phi$. The formulas for this case are given in the Appendix. The form factors for this decay have been measured rather precisely by BaBar \cite{BaBar-f0},
and we use these for further analysis of this mode. Our aim is to determine our own rate at $q^2$ equals zero, so as to cancel systematic errors in the ratio of rates for $f_0e^+\nu$/$\phi e^+\nu$.

\begin{figure}[htbp]
 \vskip 0.00cm
\centerline{\epsfxsize=6.0in \epsffile{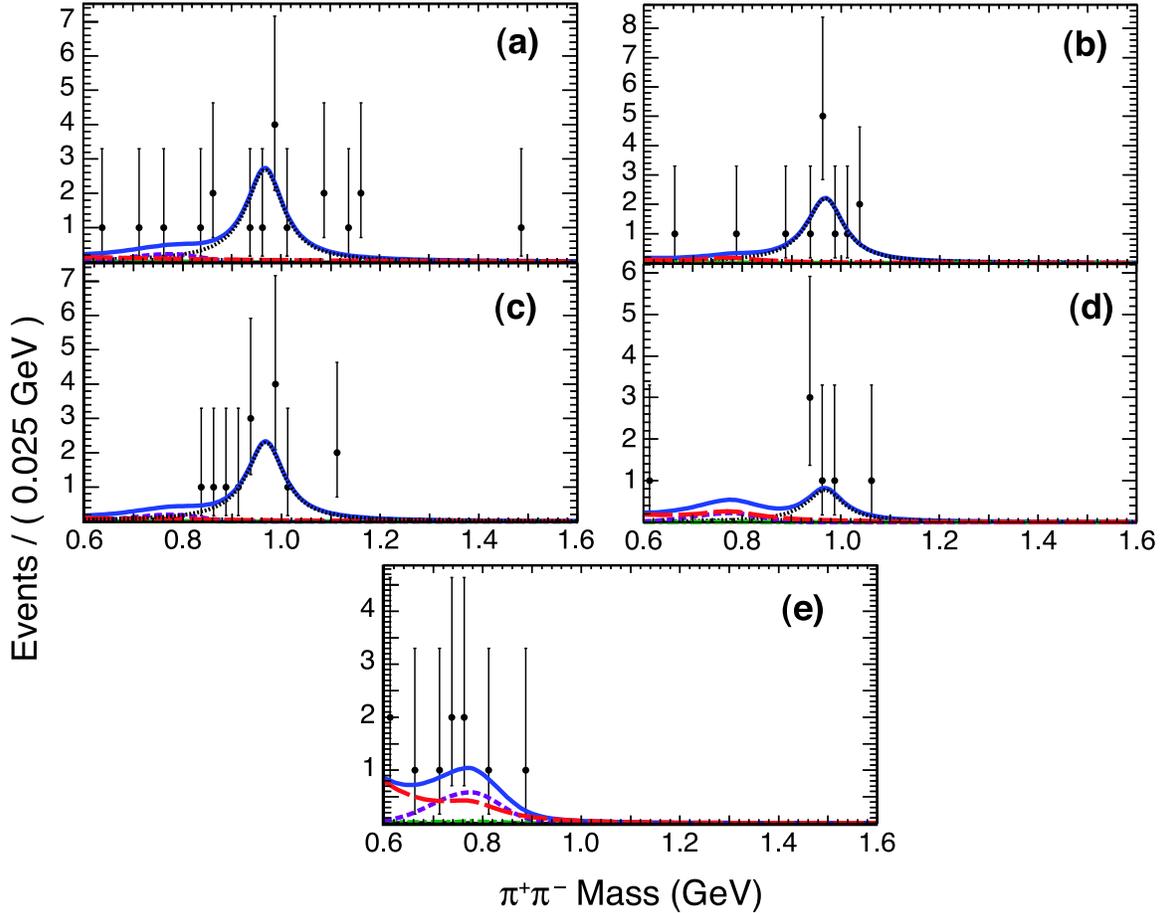}}
 \caption{The invariant  $\pi^+\pi^-$ mass in the semileptonic mode for the different $q^2$ intervals in units of GeV$^2$: (a) 0-0.2, (b) 0.2-0.4, (c) 0.4-0.6, (d) 0.6-0.8 and (e) 0.8-2.0. The fits to the data are described in the text.
 The dotted curves show the signal, the long-dashed distributions are from sidebands of the $D_s^-$ candidate invariant mass distributions, while the shorter dashed curves show the $\eta'\to\pi^+\pi^-\gamma$ background level, and the dashed-dot curves show the (small) $\pi^+\pi^-\pi^0$ background from $\phi e^+\nu$. The solid curves show the total.}
 \label{f0-qsq}
 \end{figure}

\begin{figure}[htbp]
 \vskip 0.00cm
\centerline{\epsfxsize=6.0in \epsffile{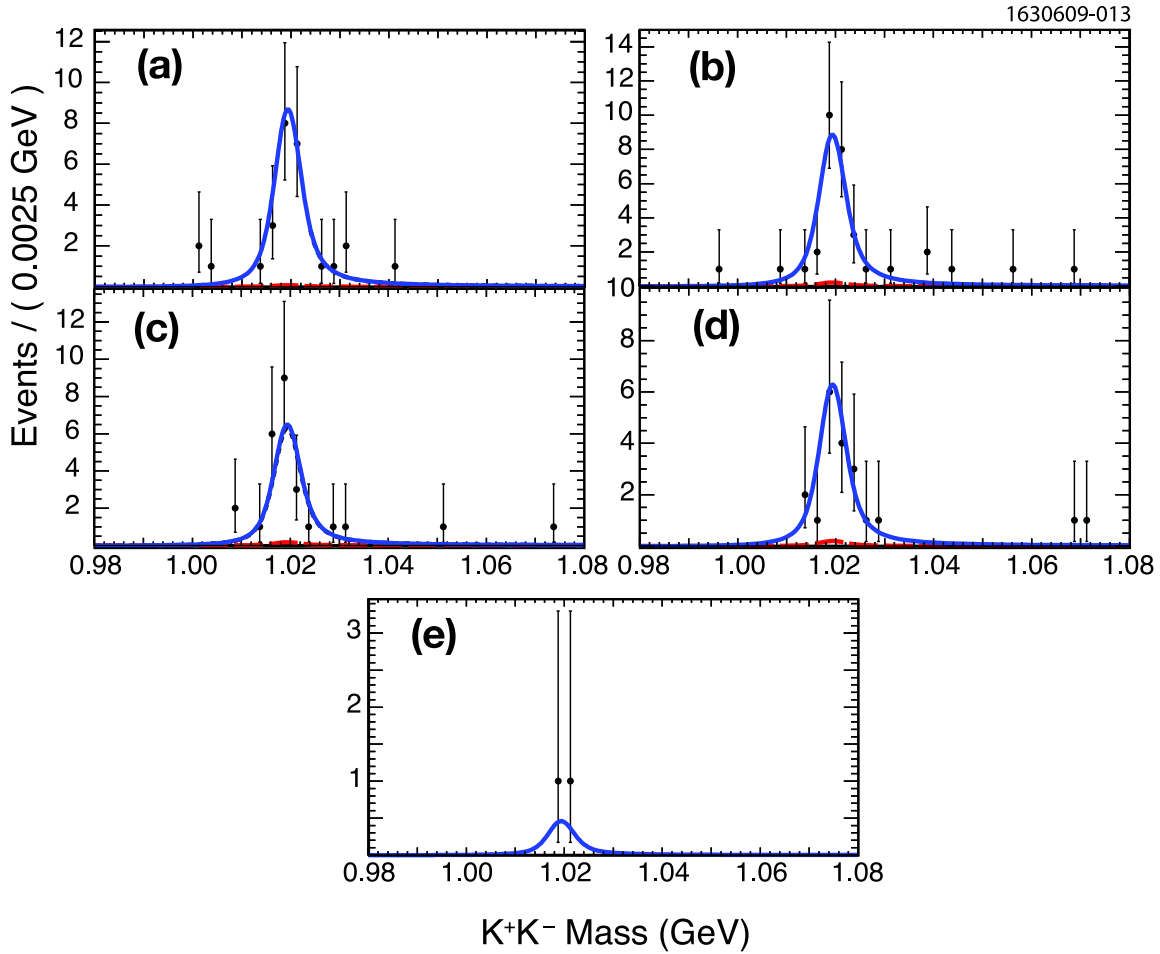}}
 \caption{The invariant  $K^+K^-$ mass in the semileptonic mode for the different $q^2$ intervals in units of GeV$^2$: (a) 0-0.2, (b) 0.2-0.4, (c) 0.4-0.6, (d) 0.6-0.8 and (e) 0.8-1.0. The solid curve show the total and
the long-dashed distributions are from sidebands of the $D_s^-$ candidate invariant mass distributions.}
 \label{phi-qsq}
 \end{figure}

We separate the data into five $q^2$ intervals and perform fits to the mass distributions. These are shown in Fig.~\ref{f0-qsq} for $\pi^+\pi^- e^+\nu$, and in Fig.~\ref{phi-qsq} for $K^+K^- e^+\nu$. The resulting signal yields are given in Table~\ref{tab:qsq}. For the $\pi^+\pi^- e^+\nu$ mode, the fitting function consists of a sum of a signal Breit-Wigner with mass and width fixed to the values we found by summing all $q^2$ intervals for the $f_0$, a fixed background function from the $D_s^-$ sidebands, a floating background shape to account for $D_s^+\to\eta' e^+\nu$, $\eta'\to\pi^+\pi^-\gamma$, and a fixed background function to account for $D_s^+\to\phi e^+\nu$, $\phi\to\pi^+\pi^-\pi^0$. The $K^+ K^- e^+\nu$ shape is defined in section~\ref{sec-IMD} and is kept the same in each $q^2$ interval.

\begin{table}[htb]
\begin{center}
\caption{Number of events (\#) and efficiencies $\epsilon$ in $q^2$ intervals. \label{tab:qsq}}
\begin{tabular}{cccccccc}
 \hline\hline
 $q^2$ interval & \#$f_0 e^+\nu$ & $\epsilon(f_0)$ &\#$f_0e^+\nu$ &  \#$\phi e^+\nu$ & $\epsilon(\phi)$ &\#$\phi e^+\nu$ \\
 (GeV$^2$)& & (\%) &corrected &&(\%)&corrected
 \\\hline
~~0-0.2 & 14.6$\pm$3.9  & 45.5 &32.1$\pm$8.6 & 30.5$\pm$5.1& 25.9& 118$\pm$20  \\
0.2-0.4 & 12.3$\pm$3.5  & 50.2 &24.5$\pm$7.0 & 30.5$\pm$5.3& 22.9& 133$\pm$23  \\
0.4-0.6 & 12.4$\pm$3.5  & 53.6 &23.1$\pm$6.5 & 22.3$\pm$4.4& 21.5& 104$\pm$20 \\
0.6-0.8 & ~4.2$\pm$2.2  & 56.9 &~7.4$\pm$3.9 & 21.5$\pm$4.7& 15.6& 138$\pm$30\\
0.8-2.0 & ~~~0$\pm$1.0  & 55.9 &~0$\pm$1.8   & ~1.6$\pm$1.1& ~9.8& ~16$\pm$11 \\\hline
Sum     & 43.5$\pm$6.7  &      &87.1$\pm$13.5&106.4$\pm$9.8&     & 508$\pm$49\\
\hline\hline
\end{tabular}
\end{center}
\end{table}

The product branching fractions can be extracted by dividing the efficiency-corrected event sums in each mode by the number of $D_s^-$ tags. We find
\begin{eqnarray}
{\cal B}(D_s^+\to f_0(980) e^+ \nu,~f_0\to \pi^+\pi^-)&=&(0.20\pm 0.03\pm 0.01)\%, \\\nonumber
{\cal B}(D_s^+\to \phi e^+ \nu,~\phi\to K^+K^-)&=&(1.16\pm 0.11 \pm 0.06)\%~,
\end{eqnarray}
where the first error is statistical and the second systematic (see section~\ref{sec:syserr}).  These values are consistent with, though the $f_0$ mode is somewhat larger than, our previously published values of
(0.13$\pm$0.04$\pm$0.01)\%\footnote{Our result is larger here because the previous analysis assumed the $f_0$ natural
 width was 50 MeV.} and (1.13$\pm$0.19$\pm$0.06)\% for the $f_0$ and $\phi$ modes, respectively. We can compare directly with the branching ratio measurement of the $\phi$ mode of BaBar \cite{BaBar-f0} by dividing by the $\phi\to K^+ K^-$ branching fraction of (49.2$\pm$0.6)\% \cite{PDG}. Our measurement then becomes
\begin{equation}
{\cal B}(D_s^+\to \phi e^+ \nu)=(2.36\pm 0.23\pm 0.13)\%~,
\end{equation}
while the BaBar measurement is ($2.61\pm0.03\pm0.08\pm0.15$)\%.

\begin{figure}[htbp]
 \vskip 0.00cm
\centerline{\epsfxsize=6.0in \epsffile{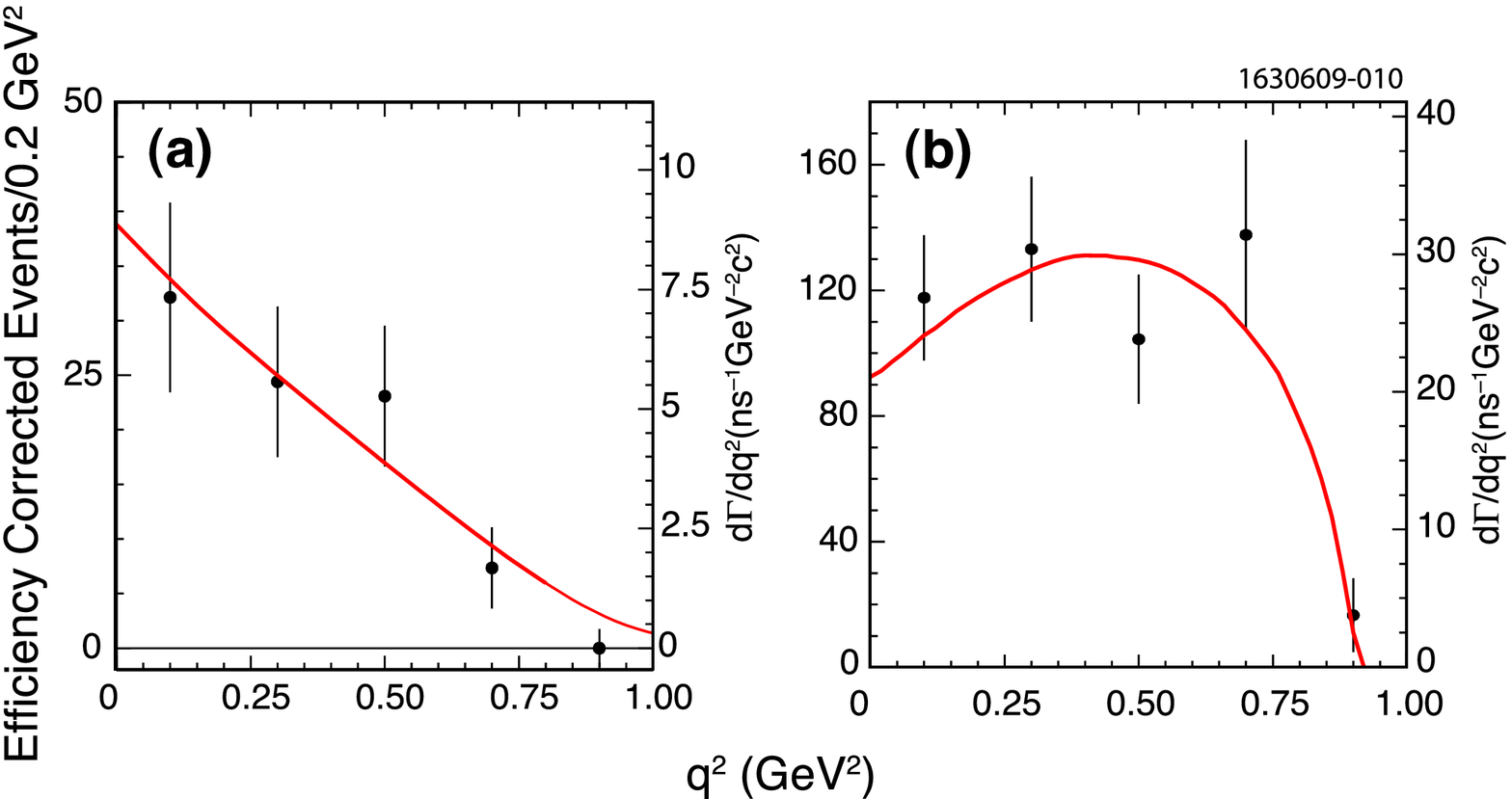}}
 \caption{The $q^2$ distributions for (a) $D_s^+\to f_0 e^+\nu$ fit to a function allowing a varying pole mass and (b) $D_s^+\to \phi e^+\nu$ fit to a function with form factors fixed to those measured by BaBar \cite{BaBar-f0}.}
 \label{fitq2}
 \end{figure}

Figure~\ref{fitq2} shows the $q^2$ distributions for the $f_0 e^+\nu$ and $\phi e^+ \nu$ channels. We fit the $\phi e^+ \nu$ channel using the form factors determined by the BaBar collaboration letting only the normalization float \cite{BaBar-f0}. The fit yields 478$\pm$45 events compared with 508$\pm$49 from summing the data.

For $f_0 e^+\nu$ we fit the $q^2$ distribution to Eq.~\ref{eq:f0q2} using a simple pole model for the shape of the form factor, $|f_+(q^2)|=1/(1-q^2/M_{\rm pole}^2)$. (We exclude the highest $q^2$ point from the fit.) This procedure gives a value of (1.7$^{+4.5}_{-0.7}$) GeV for $M_{\rm pole}$. We then vary our measured $f_0$ width by one standard deviation and repeat the procedure to find the systematic uncertainty, resulting in a $\pm$0.2 GeV systematic error. Using the central value of the pole mass we find that the fitted yield for $f_0 e^+\nu$ is 86.6$\pm$13.0 corrected events compared with 87.1$\pm$13.5 corrected events found by summing over the $q^2$ intervals.

We now use these fits to determine $R_{f/\phi}$. At zero $q^2$ we find 38.8$\pm$9.3 corrected $f_0 e^+\nu$ events, and 92.6$\pm8.7$ $\phi e^+\nu$ events. Thus
\begin{equation}
R_{f/\phi}=(42\pm11)\%.
\end{equation}
Here any errors in the event tagging or reconstruction efficiencies cancel. The systematic error due to the uncertainties on the BaBar form factors are much smaller than the statistical errors.

\subsection{Systematic Errors}
\label{sec:syserr}

The systematic errors on these branching fractions are given in
Table~\ref{tab:syserr}. The error on track finding is determined
from a detailed comparison of the simulation with double tag
events where the entire event can be reconstructed even if one track is missed  \cite{absDbr}.
The particle identification uncertainty on the
two hadronic tracks is listed as twice the uncertainty on the identification of one hadron.
The error on the number of tags of $\pm$2\% is assigned by varying the
fitting functions and ranges.
Additional systematic errors arising from
the background estimates are at the 1\% level. Final state radiation effects are included in our simulations
and the resulting accumulated uncertainty from the two-hadrons and the lepton is on the order of 1\%.

We have previously checked the resolution in MM$^2$ by fitting to a sample of $D_s^+\to K^+ K^0$ candidates where we first found the $K^0$ and then ignored it \cite{Dstomunu}. The resolution agreed with the simulation to 1.7\%. Here, because we have a positron in the final state instead of a $K^+$ we use an uncertainty of $\pm$3.4\% to allow for uncertainties associated with radiation in detector material. We have made wide selection cuts to minimize the effect of any resolution errors on our extracted rates.

We note that there is no enhancement in our ability to find
tags in $f_0 e^+\nu$ or $\phi e^+\nu$
events (tag bias) as compared with events where the $D_s^+$ decays generically.
Using a Monte Carlo simulation for each tag mode
independently and then average the results based on the known tag fractions, we find
a correction factor of (0$\pm$1\%), which we assign as a systematic error.

\begin{table}[htb]
\begin{center}
\caption{Systematic errors on determination of the branching fractions. \label{tab:syserr}}
\begin{tabular}{lc} \hline\hline
   Error Source & Size (\%) \\ \hline
Track finding &2.1 \\
Hadron identification & 2.0\\
Electron identification &1.0 \\
MM$^2$ width & 3.4\\
Background & 1.0\\
Number of tags& 2.0\\
Tag bias & 1.0\\
\hline
Total & 5.2\\
 \hline\hline
\end{tabular}
\end{center}
\end{table}

\section{Further Results and Implications}
\subsection{\boldmath Strange Mixing Angle of the $f_0(980)$}

In one class of models, the $f_0(980)$ is thought to be a $J^P=0^+$ state described in the quark model as composed of quark-antiquark states which form a mixture of strange and non-strange components characterized
by a mixing angle $\theta$ defined in terms of the wave-function as described by Aliev and Savci \cite{Aliev}
\begin{equation}
\mid f_0\rangle=\cos\theta\mid\overline{s}s\rangle+
\sin\theta\frac{1}{\sqrt{2}}\mid\left(\overline{u}u+\overline{d}d\right)\rangle~.
\end{equation}

Since the central value of the mass is below threshold for decay into a kaon pair, the branching ratio
to physical $\overline{s}s$ states is suppressed.
BES has extracted relative branching ratios using $\psi(2S)\to \gamma \chi_{c0}$ decays where the $\chi_{c0}\to f_0(980)f_0(980)$, and either both $f_0(980)$'s decay into $\pi^+\pi^-$ or one into $\pi^+\pi^-$ and the other into $K^+K^-$ \cite{BES-chi}. From their results we obtain
\begin{equation}
\frac{{\cal B}(f_0\to K^+ K^-)}{{\cal B}(f_0\to \pi^+\pi^-)}=(25^{+17}_{-11})\%~.
\end{equation}

Assuming that the $\pi\pi$ and $KK$ decays are dominant  we can also extract
\begin{equation}
{\cal B}(f_0\to \pi^+\pi^-)=(50^{+7}_{-9})\%~,
\end{equation}
where we have assumed that the only other decays are to $\pi^0\pi^0$, 1/2 of the $\pi^+\pi^-$ rate, and to neutral kaons, equal to charged kaons. This approach has been used by B. El-Bennich \etal~\cite{mixangothers}.
It is possible, however, that the decays $f_0\to \pi^+\pi^-\pi^+\pi^-$ or  $\pi^+\pi^-\pi^0\pi^0$
are significant. If so, then the branching fraction of $f_0$ into two pions will be smaller,

 Using the above branching fraction we extract
${\cal B}(D_s^+\to f_0 e^+\nu)=(0.40\pm 0.06\pm 0.06)\%$, where the error on the $f_0\to\pi^+\pi^-$ rate has been included in the systematic error, and, in fact, is the dominant contribution.

We can use this rate to estimate
the mixing angle $\cos\theta$. Aliev and Savci using QCD
sum rules predict that ${\cal B}(D_s^+\to f_0 e^+\nu)=\cos^2\theta\times(0.41)\%$. This
provides us with a value of
\begin{equation}
\cos^2\theta=0.98^{+0.02}_{-0.21}~,
\end{equation}
where we have truncated the positive uncertainly so that the value does not exceed unity.
This determination gives a somewhat larger but consistent evaluation with other methods \cite{mixangothers}.
If the $f_0$ decay rate into four pions is significant, our value of the mixing angle will be somewhat larger.

The near unity value for $\cos^2\theta$ poses a conundrum: If the $f_0$ is dominantly an $\mid\overline{s}s\rangle$
state, why does it decay predominantly into $\pi\pi$? A solution may be that the quark content of the $f_0$ is not composed of just a quark-antiquark state, but may contain a some four-quark content \cite{Joe}.

\subsection{\boldmath The $f_0(980)$ Mass and Width}

We now return to the determination of the $f_0$ mass and width. In the fit described above we
used a relativistic Breit-Wigner function. We did not, however, take into account that the phase space
for the larger $f_0$ masses is somewhat smaller, due to the finite $D_s$ mass, than for smaller $f_0$ masses.
The changes to the invariant mass spectrum of the $f_0$ have been parameterized by Dosch \etal~\cite{DFNN}.
The double differential decay rate in terms of $s\equiv M^2_{\pi\pi}$ is
\begin{equation}
\frac{d^2\Gamma(s,q^2)}{dsdq^2}=\frac{G_F^2|V_{cs}|^2}{192\pi^4M_{D_s}^3}
\lambda^{3/2}(M_{D_s}^2,s,q^2)|f^+(q^2)|^2 P(s),
\end{equation}
where $\lambda(x,y,z)=x^2+y^2+z^2-2xy-2xz-2yz$ and is equal to $4M_{D_s}^2$ times the square of the momentum of the $f_0$ in the $D_s$ rest frame, and  $P(s)$ is the relativistic Breit-Wigner shape given in equations~\ref{eq:relBW} and \ref{eq:relBW2}.
The resulting, slightly shifted mass spectrum, then is given by integrating the above expression over $q^2$
\begin{equation}
\frac{d\Gamma}{d\sqrt{s}}=2\sqrt{s}\int_0^{\left(M_{D_s}-\sqrt{s}\right)^2} \frac{d^2\Gamma(s,q^2)}{dsdq^2} dq^2~.
\end{equation}

\begin{figure}[htbp]
 \vskip 0.00cm
\centerline{\epsfxsize=6.0in \epsffile{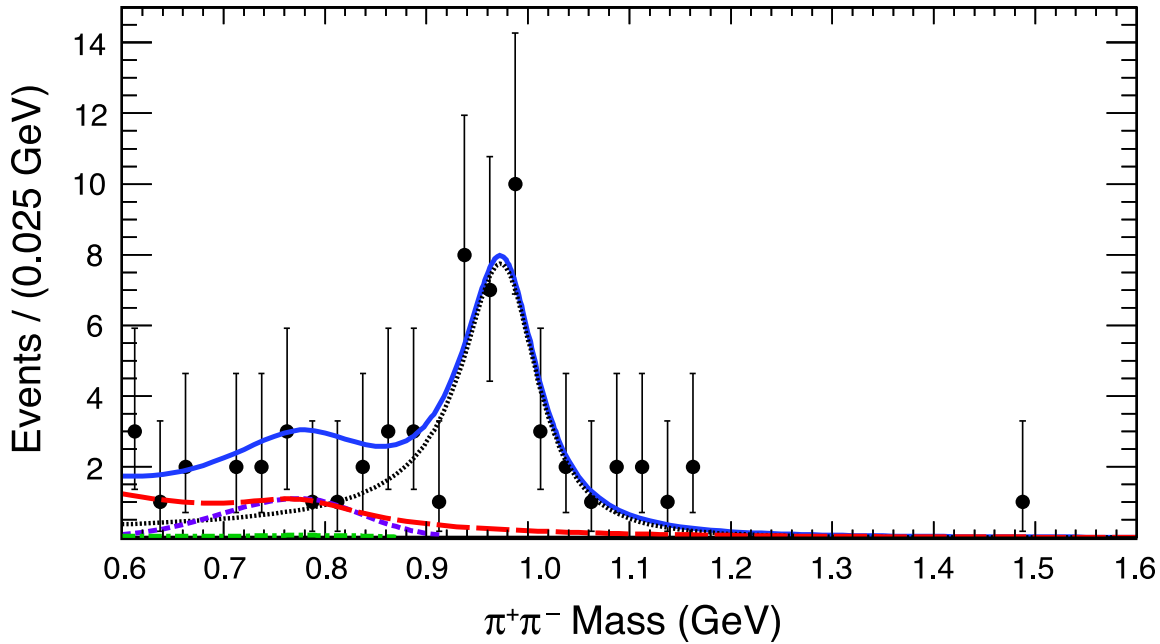}}
 \caption{Invariant mass distribution for $\pi^+\pi^-$ in the semileptonic mode.
 The dotted curve shows the signal shape, a relativistic Breit-Wigner modified by phase space and form factor effects, the long-dashed distributions are from sidebands of the $D_s^-$ candidate invariant mass distributions while the shorter dashed curve in  shows the $\eta'\to\pi^+\pi^-\gamma$ background level, and the dashed-dot curve shows the (small) $\pi^+\pi^-\pi^0$ background from $\phi e^+\nu$. The solid curve shows the total. }
 \label{f0mm_new}
 \end{figure}

We perform this integral using our fitted form factor shape for $|f_+(q^2)|$ and our original fitted parameters
for the relativistic Breit-Wigner function. This generates a new shape for the mass spectrum, that we fit to the data.
In principle this procedure can be iterated many times; here we find that only one iteration is sufficient.
The resulting curve fitted to the data is shown in Fig.~\ref{f0mm_new}. The systematic errors on the mass
are small compared to the statistical error. The masses of known particles, $K_S$, $\Lambda$, $\phi$, $D^0$ are reproduced in other CLEO analyses to better than 1 MeV. Changing the background level by $\pm$1 standard deviation changes the mass by a negligible amount. For the width we have identified two sources of systematic error, one the resonance shape and the other the form factor shape. Changing the Breit-Wigner shape from relativistic to non-relativistic causes a change of $\pm$2 MeV. We also vary the form factor by $\pm$ one standard deviation resulting in a change of about $\pm$2 MeV. Our final values for the mass and width are
\begin{equation}
M_{f_0}=(977^{+11}_{-9}\pm 1){\rm~MeV},~~~~{\rm and~} \Gamma_0=(91^{+30}_{-22}\pm 3)~{\rm MeV}~.
\end{equation}
We do not fit our data to a function that allows for the opening of a threshold in another channel, because
there is no obvious change in the shape of the dipion mass spectrum near 1 GeV, and we do not have enough data
to specify the parameters of such a fit.

\section{Conclusions}

We present an updated result for the first measurement of
\begin{equation}
{\cal B}(D_s^+\to f_0(980) e^+ \nu){\cal B}(f_0\to \pi^+\pi^-)=(0.20\pm 0.03\pm 0.01)\%.
\end{equation}
Assuming a simple pole model for the form factor, $|f_+(q^2)|$, we estimate the pole mass as
($1.7^{+4.5}_{-0.7}\pm 0.2$) GeV.

In the final state the only hadron present is the $f_0$. This provides a particulary clean environment that allows us to measure the mass and width as $(977^{+11}_{-9}\pm 1){\rm~MeV}$, and $(91^{+30}_{-22}\pm 3)~{\rm MeV}$, respectively.

We update our measurement of
\begin{equation}
{\cal B}(D_s^+\to \phi e^+ \nu)=(2.36\pm 0.23\pm 0.13)\%~.
\end{equation}

We measure the ratio of decay rates at $q^2=0$
\begin{equation}
R_{f/\phi}\equiv \frac{{\frac{d\Gamma}{dq^2}}(D_s^+\to f_0(980) e^+\nu,~f_0\to\pi^+\pi^-)\mid_{q^2=0}}{{\frac{d\Gamma}{dq^2}}(D_s^+\to \phi e^+\nu,~\phi\to K^+ K^-)\mid_{q^2=0}}=(42\pm11)\%~.
\end{equation}
This ratio has been predicted by Stone and Zhang to equal \cite{SZ}
\begin{equation}
R_{f/\phi}=\frac{\Gamma(B_s\to J/\psi f_0,~f_0\to \pi^+\pi^-)}{\Gamma(B_s\to J/\psi\phi,~\phi\to K^+K^-)}.
\end{equation}
Our measurement indicates that the $B_s\to J/\psi f_0$ channel may indeed be a useful place to measure CP violation in the $B_s$ system in that the rate can be $\sim$40\% that for $J/\psi \phi$ mode, especially since an angular analysis is not necessary, as the $J/\psi f_0$ mode is a CP eigenstate.

Finally, we show that viewing the $f_0$ as a quark-antiquark state results in a mixing angle between $\overline{s}{s}$ and light quarks whose cosine is close to unity, consistent with other estimates \cite{mixangothers}. Thus, the large rate into $\pi\pi$ may be the result of part of the wave-function being a four-quark state \cite{Joe}.

\section{Acknowledgments}
We thank Joe Schechter and Justine Serrano for useful conversations.
We gratefully acknowledge the effort of the CESR staff
in providing us with excellent luminosity and running conditions.
D.~Cronin-Hennessy and A.~Ryd thank the A.P.~Sloan Foundation.
This work was supported by the National Science Foundation,
the U.S. Department of Energy,
the Natural Sciences and Engineering Research Council of Canada, and
the U.K. Science and Technology Facilities Council.

\section*{APPENDIX}
We briefly summarize the relationships between different rates and form factors in $D_s$ decays into a vector particle ($\phi$) here. There are more Lorentz invariants possible than in the pseudoscalar to scalar case, and
three independent form factors, $V$, $A_1$, and $A_2$ are involved in the case of $e^+\nu$ decays. The matrix element can be written as the sum of a vector plus axial-vector current \cite{Artuso-semi}:
\begin{eqnarray}
\langle \phi | \bar s \gamma_\mu c | D_s \rangle &=&
2\frac{V(q^2)}{M_{D_s}+M_{\phi}} \epsilon_{\mu\nu\alpha\beta}
p_{D_s}^\nu p_{\phi}^\alpha \epsilon^{*\beta},
~~{\rm and} \\
\langle \phi | \bar s \gamma_\mu \gamma_5 c | D_s \rangle &=&
i \left(M_{D_s} + M_{\phi}\right)
\left(\epsilon^*_\mu - \frac{\epsilon^*\cdot q}{q^2} q_\mu \right) A_1(q^2)
\\
&-& i \frac{\epsilon^*\cdot q}{M_{D_s}+M_{\phi}}
\left(P^\mu-\frac{M_{D_s}^2-M_{\phi}^2}{q^2} q^\mu \right) A_2(q^2)
\nonumber
\end{eqnarray}
where $\epsilon^*$ is a polarization of the final state meson. (An additional term proportional to
$A_0$ has been dropped as it gets multiplied by the square of the electron mass.)

First of all the differential decay rate can be written in terms of the helicity amplitudes as \cite{GS}
\begin{equation}
\frac{d\Gamma}{dq^2}=\frac{G_F^2|V_{cs}|^2}{96\pi^3}\frac{p_{\phi}q^2}{M^2_{D_s}}\left(|H_+|^2+|H_-|^2+|H_0|^2\right)~.
\end{equation}
The helicity amplitudes are expressed in terms of three form factors in the limit of zero lepton mass as \cite{hff}
\begin{eqnarray}
H_0(q^2)&=&\frac{1}{2M_{\phi}\sqrt{q^2}}\left[\left(M_{D_s}^2-M_{\phi}^2-q^2\right)(M_{D_s}+M_{\phi})A_1(q^2)-4\frac{M_{Ds}^2 p_{\phi}^2}{M_{Ds}+M_{\phi}}A_2(q^2)\right],\\
H_{\pm}(q^2)&=&(M_{D_s}+M_{\phi})A_1(q^2)\mp \frac{2M_{D_s}p_{\phi}}{(M_{D_s}+M_{\phi})}V(q^2)~.
\end{eqnarray}

The most elementary parameterization of the form factors considers the $q^2$ dependence as the behavior of a simple pole in the $t$-channel, yielding
\begin{equation}
V(q^2)=\frac{V(0)}{1-q^2/M_V^2};~~~~A_i(q^2)=\frac{A_i(0)}{1-q^2/M_A^2},~{\rm for}~i=1~{\rm and}~2.
\end{equation}
The masses can be taken as $M_V=M_{D_{s}^*}$, and $M_A=M_{D_{s1}}$, although they can, in principle, be extracted from the data.
The BaBar collaboration has assumed the value for $M_V$ and measured $M_A=\left(2.28^{+0.23}_{-0.18}\pm 0.18\right)$ GeV; they also determined at $q^2=0$, $r_V = V(0)/A_1(0) = 1.849\pm 0.060\pm 0.095;$ $r_2 = A_2(0)/A_1(0) = 0.763\pm 0.071\pm 0.065$ \cite{BaBar-f0}.

\afterpage{\clearpage}

\end{document}